# Non-coalescence and in-plane momentum generation in sessile droplet clusters

Gopal Chandra Pal[†], Cheuk Wing Edmond Lam[‡], Chander Shekhar Sharma[†]*

[†]Thermofluidics Research Laboratory, Department of Mechanical Engineering, Indian Institute of Technology Ropar, Rupnagar, Punjab 140 001, India
[‡]Department of Mechanical Engineering, Massachusetts Institute of Technology, Cambridge, Massachusetts 02139, USA
*Email: chander.sharma@iitrpr.ac.in, Ph: +91-1881-232358

**Abstract**

Intuitively, droplets in proximity merge when brought into contact. However, under certain conditions, they may not coalesce due to the entrapment of an interstitial gas film. Non-coalescence between water droplets has so far been observed during collisions of droplets moving with relative centroidal velocity, or in the presence of specific enabling effects such as high intervening gas pressures, surfactants, or large droplet sizes (diameter $\gtrapprox$ 1 mm). Here, we report non-coalescence between water droplets over a much wider range of droplet diameters, from millimeters to as small as 100 microns, without the need for any of the above factors. Such non-coalescence occurs in sessile droplet clusters on water-repellent surfaces. When any two droplets in a cluster coalesce, the evolving interface of the coalescing droplets comes in apparent contact with other neighbouring droplets in the cluster, but does not necessarily trigger further coalescence. In fact, such apparent contact can manifest as a bouncing interaction, and depending on the initial geometric arrangement of droplets, it can result in significant lateral momentum generation, consequently leading to spontaneous in-plane self-propulsion of the participating droplets. The energy conversion efficiency of this process reaches as high as 9% for closely packed clusters of 3 sessile droplets and increases further with an increase in the number of participating droplets. The resulting self-propulsion of such small droplets reveals a new pathway for passive droplet removal and surface renewal during dropwise condensation on superhydrophobic surfaces, critical in multiple applications.

## Main

Miscible droplets in close proximity coalesce into a single droplet as a result of the van der Waals force of attraction between the interfaces[1]. However, in certain instances, droplets bounce from each other instead of merging due to the retention of a thin gaseous film, typically ~10-100 nm thick,[2] between them. The non-coalescence (bounce) of droplets has been observed in multiple scenarios, such as rain droplets in air,[3–5] oppositely charged droplets[6], droplets in the Leidenfrost state,[7,8] droplets on a pool of liquid,[9,10] and colliding droplets.[11,12] The stable gas film formed between the interfaces of such non-coalescing droplets can consist of either air[13] or the vapor generated from the droplet through evaporation.[7] The stability of this interstitial gaseous film can depend on the relative velocity[12–14] and the relative position of droplets,[15] and residence time (the time duration for which the interfaces of droplets stay in apparent contact prior to coalescence).[16] Despite such widespread observations of non-coalescence between droplets, non-coalescence for water droplets has hitherto been reported only for large, millimeter-sized droplets (diameter > 2 mm) colliding at high velocities while moving in air or on water repellant surfaces.[12,17–20]

This article reports for the first time the non-coalescence of water droplets over a large range of droplet sizes, ranging from millimetric droplets to droplets as small as 100 µm in diameter, and with no bulk velocity prior to coming in apparent contact. This non-coalescence is observed in sessile droplet clusters on water-repellant surfaces. Further, this non-coalescence manifests without the presence of any external fields, surfactants, and any gradients in fluid properties. Specifically, we find that when two droplets in a droplet cluster start to coalesce, the resulting capillary ripple does not necessarily trigger further coalescence with other droplets in the cluster. Figure 1 illustrates an exemplary case of such non-coalescence. Here, coalescence is initiated between two millimetric sessile droplets $D_1$ and $D_2$ and their interface subsequently evolves and comes into apparent contact with the third sessile droplet $D_3$. However, instead of triggering coalescence with $D_3$, the interface bounces against droplet $D_3$ and retracts.

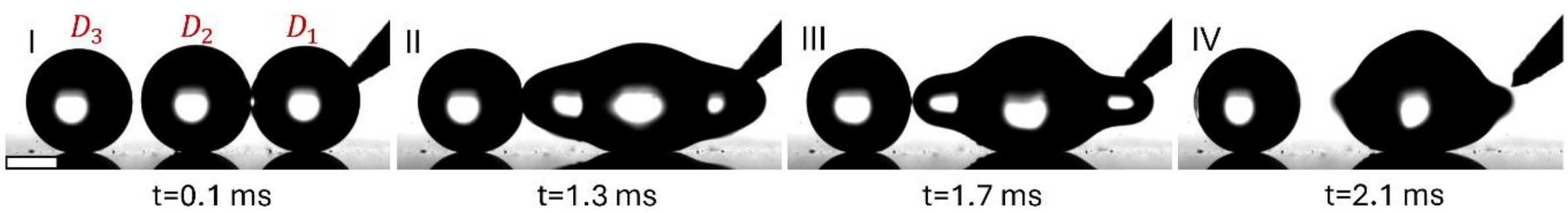

***Figure 1**: Non-coalescence of sessile droplet clusters. The scale bar represents 0.5 mm. The image sequence shows the initiation of coalescence between droplets $D_1$ and $D_2$* (frame I), *the time instants at which the interface of the merged droplet $D_1 + D_2$ comes in apparent contact with the interface of the droplet $D_3$* (frames II and III), *and subsequent retraction of $D_1 + D_2$* interface without coalescence (IV).

We further observe that when any two droplets in a cluster coalesce, the outcome for the rest of the droplets in the cluster depends on their relative geometric arrangement prior to the event. The morphology and the dynamics of the interstitial gas film between droplets depend on this arrangement, and it governs the nature of apparent contact between the droplets. Non-coalescence of cluster droplets can also generate significant lateral momentum and consequently, self-propulsion of the non-coalescing droplets across the surface. This phenomenon also manifests for small condensate droplets, thus revealing a new pathway for spontaneous droplet removal from the surface, critical for such applications.

**Results and discussion**

We investigate interactions among sessile droplets on nanotextured superhydrophobic substrates. We prepare these substrates either by spray coating silanized silica nanoparticles dispersed in isopropanol[21–23] on glass slides, or by silanizing aluminum substrate nanotextured using the Boehmite process[24]. We carefully position the droplets in clusters with different geometric arrangements on these substrates and initiate coalescence between two droplets by using a superhydrophobic tip[25] (Refer to Materials and Methods and Section S1 for further details). The experiments are performed for droplet clusters consisting of similarly sized droplets, with nominal droplet diameters, $d_0$ ranging between 0.6 and 2 mm.

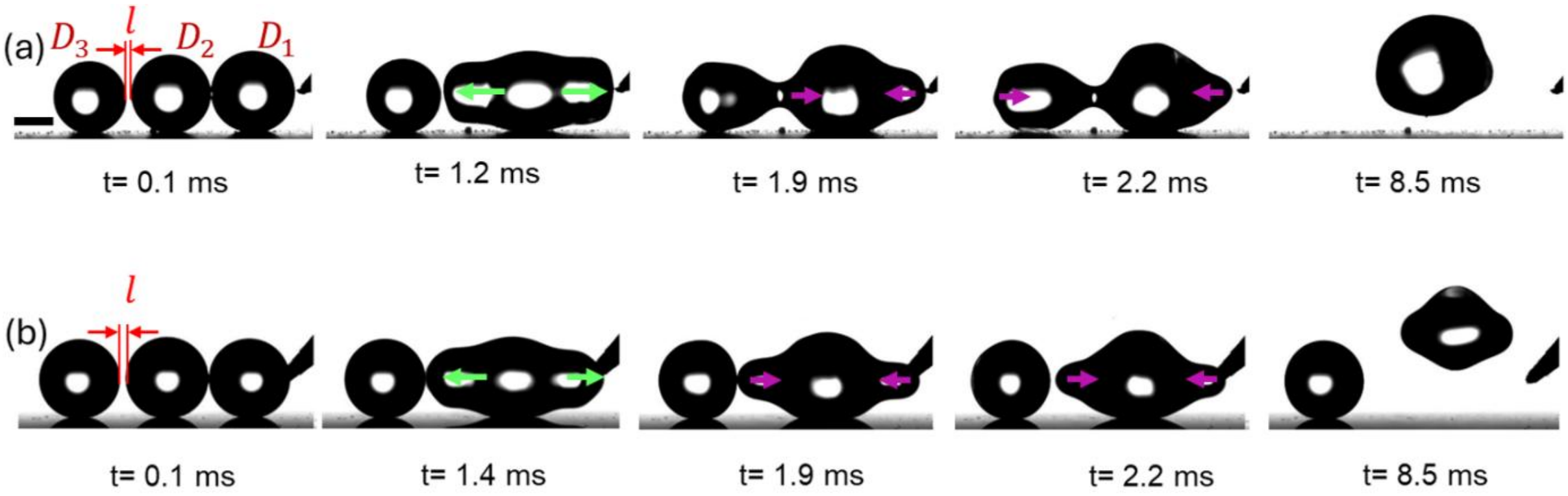

***Figure 2**: Coalescence of a droplet cluster consisting of three droplets in the in-line arrangement. Based on the initial gap l (measured as shown in the first frame of both image sequences), the evolving interface of the merged droplet $D_1$+$D_2$ comes in apparent contact with $D_3$ and either (a) triggers coalescence with $D_3$ or (b) retracts without coalescence. Green and purple arrows indicate the advancement and retraction phase in the vertical plane, respectively. The scale bar in the figure is 0.5mm, as shown in the first frame.*

Figure 2 illustrates a droplet cluster wherein three droplets of similar size ($d_0 = 1\, mm \pm 3\%$) are positioned in an in-line arrangement wherein coalescence is triggered between droplets $D_1$ and $D_2$ and the initial gap ($l$) between droplets $D_2$ and $D_3$ is varied in the range $0.04 \lesssim {}^{l}/_{d_0} \lesssim 0.15$. Surprisingly, we observe non-coalescence between the evolving droplet $D_1$+$D_2$ and the droplet $D_3$ for most $l/_{d_0}$ ratios, with coalescence observed only in a few limiting cases. Figure 2(a) illustrates a typical non-coalescence event. Coalescence occurs only for a limited set of cluster arrangements, either for a small initial gap $\left(l/_{d_0} \lesssim 0.07\right)$ or for nearly the maximum feasible gap such that the interface of $D_1$+$D_2$ can still reach that of $D_3$ $\left(l/_{d_0} \gtrsim 0.14\right)$. While in the former case, the coalescence proceeds to completion, as shown in Fig 2(b), in the latter case, coalescence is followed by reseperation (Refer to Section S2). It is found that these thresholds for droplet coalescence and non-coalescence do not change with a change in droplet size $d_0$ as long as the droplets are positioned in in-line arrangement.

Droplets coalesce when the air film between them reduces to a thickness of tens of nanometers where van der Waals force of attraction is significant.[26,27] The drainage of this interstitial air film depends on the relative velocity,[8,28] shape[12] and residence time[29] for which the two interfaces are in apparent contact. For the case of sessile droplet clusters, the aforesaid parameters for the interstitial air film between the evolving interfaces of the merging droplets and the other neighbouring droplets depend on the initial geometric arrangement of droplets in the cluster. For instance, for the in-line arrangement of droplets, these parameters depend on the $l/_{d_0}$ ratio. When the initial geometric arrangement ultimately results in delayed drainage of this air film, non-coalescence manifests. In literature, non-coalescence of individual droplets has been studied mainly for collisions between droplets moving with finite bulk velocity,[2,4,5,8,30,31] without resolving details about the air drainage dynamics due to the inherent small length and time scales involved.

In order to understand the physics underlying the unexpected non-coalescence of stationary droplets observed here, we investigate the dynamics of this air layer through direct numerical simulations.[30,32,33] We focus on non-coalescence in a sessile droplet cluster with in-line arrangement. The computational domain is chosen as shown in Fig. 3a due to the inherent symmetry of the process. (Refer to Methods and Section S3 for details).[34] The two-phase simulation begins at the initiation of coalescence between $D_1$ and $D_2$ and solves for temporal evolution of droplet morphologies and flow fields in the cluster. As coalescence between $D_1$ and $D_2$ proceeds and the coalescence neck expands, a capillary ripple travels away from the neck along the interface of the coalescing droplets. This causes the interface of the merging droplet $(D_1 + D_2)$ to approach and come in apparent contact with the interface of $D_3$. The numerical results in Fig. 3b clearly demonstrate that this interface movement causes a rise in overpressure in the interstitial air film (see Fig. 3b Frame II) and subsequently the development of a region with nearly constant air film thickness between the interfaces (Frame III of Fig 3b).

Here, time, $t$, overpressure, $P$, and velocity $v$ are non-dimensionalised with inertial-capillary time scale $\tau = \sqrt{\frac{\rho d_o^3}{8\sigma}}$, capillary pressure, $P_c = \frac{8\sigma}{d_o}$ , and inertial-capillary velocity, $v_{ic} = \frac{d_0}{\tau}$ respectively.[21,35] Here, $\rho$ and $\sigma$ represent the water density and surface tension respectively. The velocity vectors show that the development of this nearly flat air film region is also accompanied by the creation of a stagnant zone within the air film due to the change in the local morphology of the interfaces.

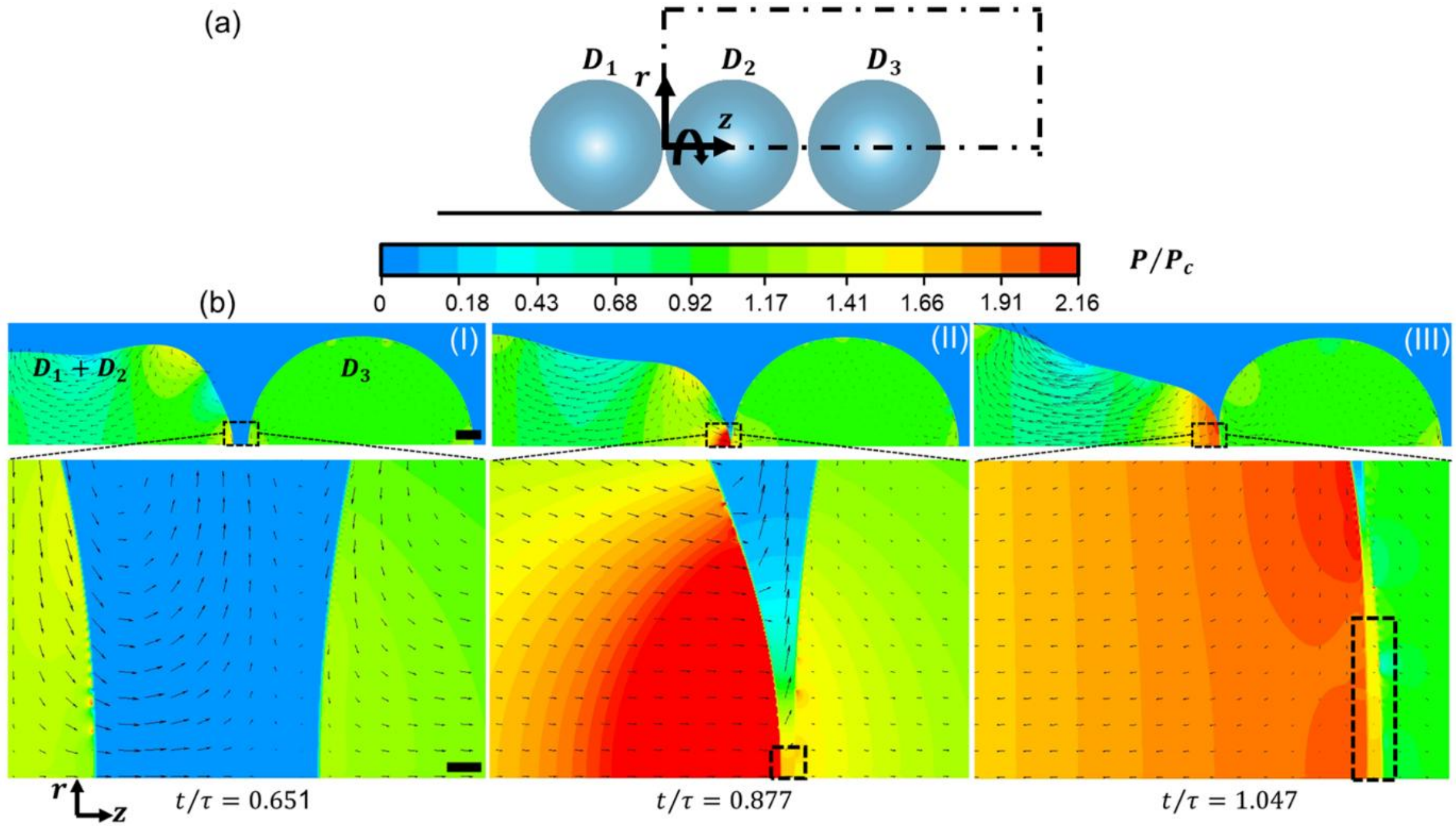

***Figure 3**: (a) The schematic shows a sessile droplet cluster in an in-line arrangement. The dashed rectangle indicates the domain of numerical simulation. A cylindrical coordinate system (r,z) is chosen with the origin at the point of coalescence between droplets $D_1$ and $D_2$. Symmetry boundary condition is imposed along the r-axis. The z-axis is chosen as the axis of rotational symmetry. (b) Image sequence illustrating the evolution of pressure and velocity fields during the event. Magnified details of the interstitial air film morphology, and pressure and velocity fields are also shown. Frame (I) shows the initial approach of $D_1$+$D_2$ interface towards $D_3$ with negligible overpressure in the interstitial air film. The overpressure in the air film rises as the interfaces come into apparent contact during the advancement phase (Frame II). Subsequently, a nearly flat air film region is formed, as marked by the dashed rectangle (Frame III). The scale bars in top and bottom image sequence represent $0.1d_0$ and $0.01d_0$, respectively. The time instants are measured from the initiation of coalescence between $D_1$* and *$D_2$.*

This evolution of the flat air film region becomes more evident from Fig. 4a. It starts to form at $t/\tau \approx 0.85$, immediately after the two interfaces come into apparent contact. As highlighted by the dashed red rectangle, the flat air film region expands radially as the interface of merged droplet $D_1$+$D_2$ continues to evolve. Eventually, at $t/\tau \approx 0.99$, the relative shape of the two interfaces results in the formation of a kink where the air film thickness is locally minimum.[32,33] This kink acts as a restriction against drainage of air from the interstitial region between the two interfaces, thus avoiding coalescence.[36,37] Subsequently, the retraction phase begins at $t/\tau \approx 1.11$, and the kink and the flat region gradually diminish.

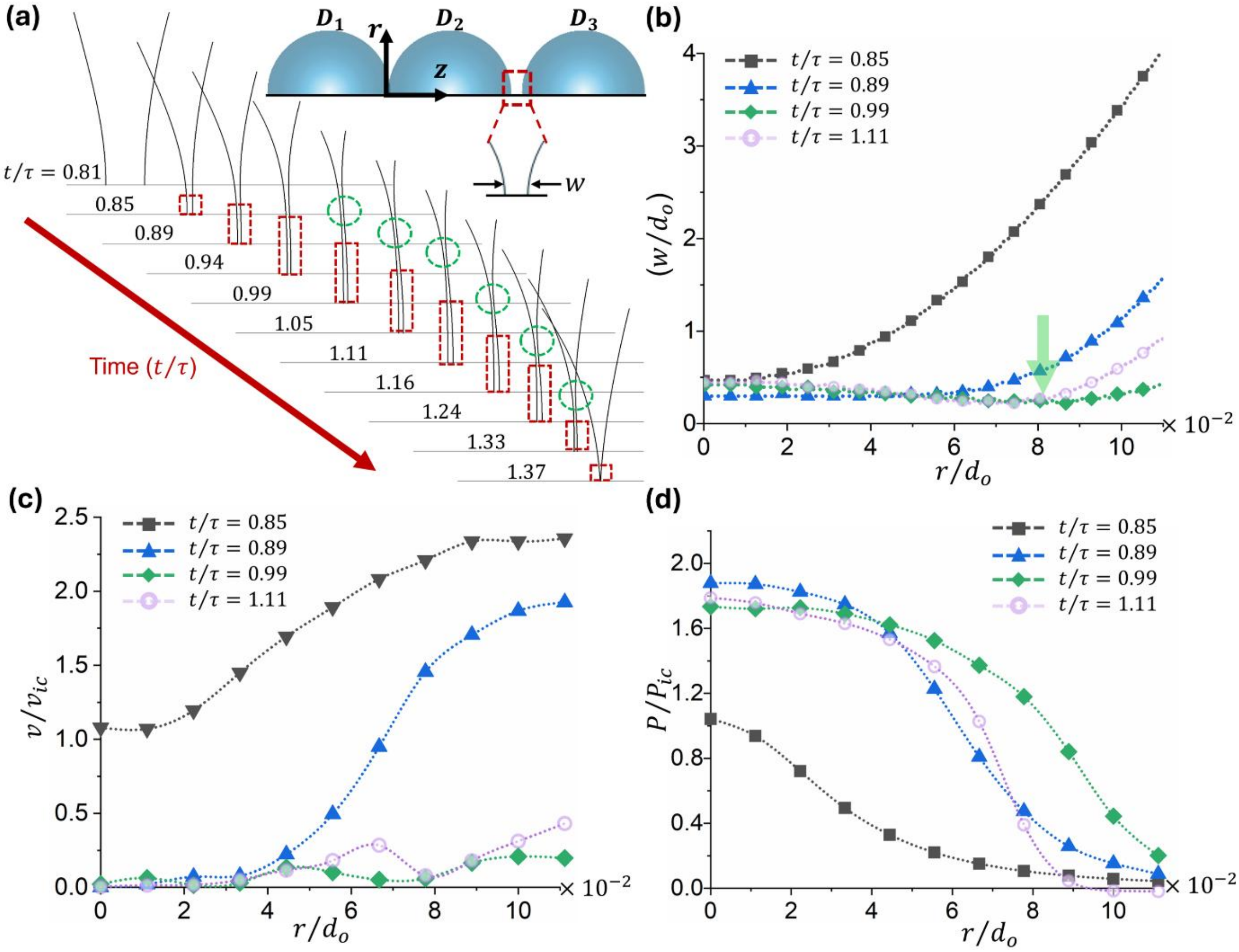

***Figure 4:*** *(a) Image sequence depicts the evolution of the interface profiles at small r, i.e. in the vicinity of z-axis (as marked by the highlighted region in inset) for non-coalescence in cluster with in-line droplet arrangement. The development of a flat air film region and the formation of a kink are highlighted by dashed red rectangles and green ovals, respectively. As the interface of merging droplets advances, the flat air film region (dashed red rectangles) develops and a kink is formed at* $t/\tau \approx 0.99$*. Subsequently, the flat air film region and the kink gradually diminish as the interface retracts. (b) Variation in the air film width* $w$ *(defined in 4(a) inset) along the radial direction during the advancement (solid symbols) and retraction phases (open symbols). Here,* $w$ *is non-dimensionalized with droplet diameter* $d_0$*. Region of nearly constant* $w/d_0$ *v/s* $r/d_0$ *represents flat air film region (for instance for* $r/d_0 \lesssim 0.05$ *at* $t/\tau \approx 0.89$*). A local minimum in* $w/d_0$ *profile at* $t/\tau \approx 0.99$ *(marked by arrow) corresponds to the formation of kink. Radial variation of non-dimensional (c) air drainage velocity* $v/v_{ic}$ *and (d) overpressure* $P/P_{ic}$ *in the interstitial air film illustrate reduction in air drainage velocity and rise in overpressure in the stagnation zone created by the kink .*

Figure 4(b) illustrates the air film thickness profiles along the radial direction at various time instants. The kink can be clearly observed at $r/d_o \approx 0.08$ for the profile corresponding to $t/\tau \approx 0.99$. The formation of the kink temporarily suppresses the drainage of air from the interstitial air film as evident from the radial profiles for relative air drainage velocity shown in Fig. 4(c), suggesting the stabilization and entrapment of air in the flat film region. Here the air drainage velocity is calculated relative to the interface of merged droplet $D_1 + D_2$ (Refer to Methods for further detail). The reduction in air drainage velocity is coupled with the rise in overpressure in the air film as shown by the radial pressure profiles in Fig. 4(d). Overall, we conclude that the kink formation leads to momentary entrapment of the air layer, leading to non-coalescence with the third droplet. As the interfaces start to react, the air drainage velocity recovers and overpressure reduces.

The morphology of the interstitial air film and its drainage dynamics depend on the local shape and relative velocity of the two interfaces. For sessile droplet clusters, these parameters are determined by the configuration of the cluster, i.e. relative geometrical arrangement of droplets prior to the event. We explore this aspect for a wide range of droplet cluster configurations by systematically varying the position of the droplet $D_3$ in the cluster, relative to droplets $D_1$ and $D_2$. In order to guide our experiments, we first estimate a radial region around $D_1$ and $D_2$ where $D_3$ can be located such that the evolving interface of the merged droplet $D_1 + D_2$ can reach $D_3$ (refer to Section S4 for further details). Subsequently, we perform experiments by varying the $D_3$ position in a radial coordinate system, with the coordinates $R$ and $\theta$ as defined in the inset of Fig. 5a. In this coordinate system, the cluster configuration with $\theta = 0$ and $1.5 \lesssim R/d_0 \lesssim 1.6$ represents the in-line arrangement of droplets discussed so far. Other configurations with $0° < \theta \leq 90°$ represent a staggered arrangement of droplets.

We compile the results of these experiments in Fig 5a and 5b. Surprisingly, non-coalescence can be observed across a wide range of cluster configurations. In contrast, full coalescence, and coalescence followed by reseparation, manifest in far fewer cases. Interestingly, we also find that for clusters with more closely packed configurations, consisting of a staggered droplet arrangement and droplet $D_3$ located in close vicinity to merging droplets $D_1$ and $D_2$, non-coalescence results in a significant generation of lateral momentum for the participating droplets. Close proximity of $D_3$ in such clusters results in large deformation of the droplet during apparent contact. The resulting increase in its surface energy acts like a compressed spring, which eventually propels the droplet along the surface with significant bulk translational kinetic energy post loss of contact with the merged droplet $D_1 + D_2$. Fig. 5c illustrates one such event. This result is in sharp contrast to the non-coalescence event in clusters with in-line configurations shown in Fig. 2b, wherein deformation of droplet $D_3$ during the event is insufficient to promote pronounced lateral movement. Fig. 5d compares in-plane displacement of droplets for the two non-coalescence event wherein non-coalescence in staggered cluster launches the droplets with significant in-plane motion compared to in-line arrangement.

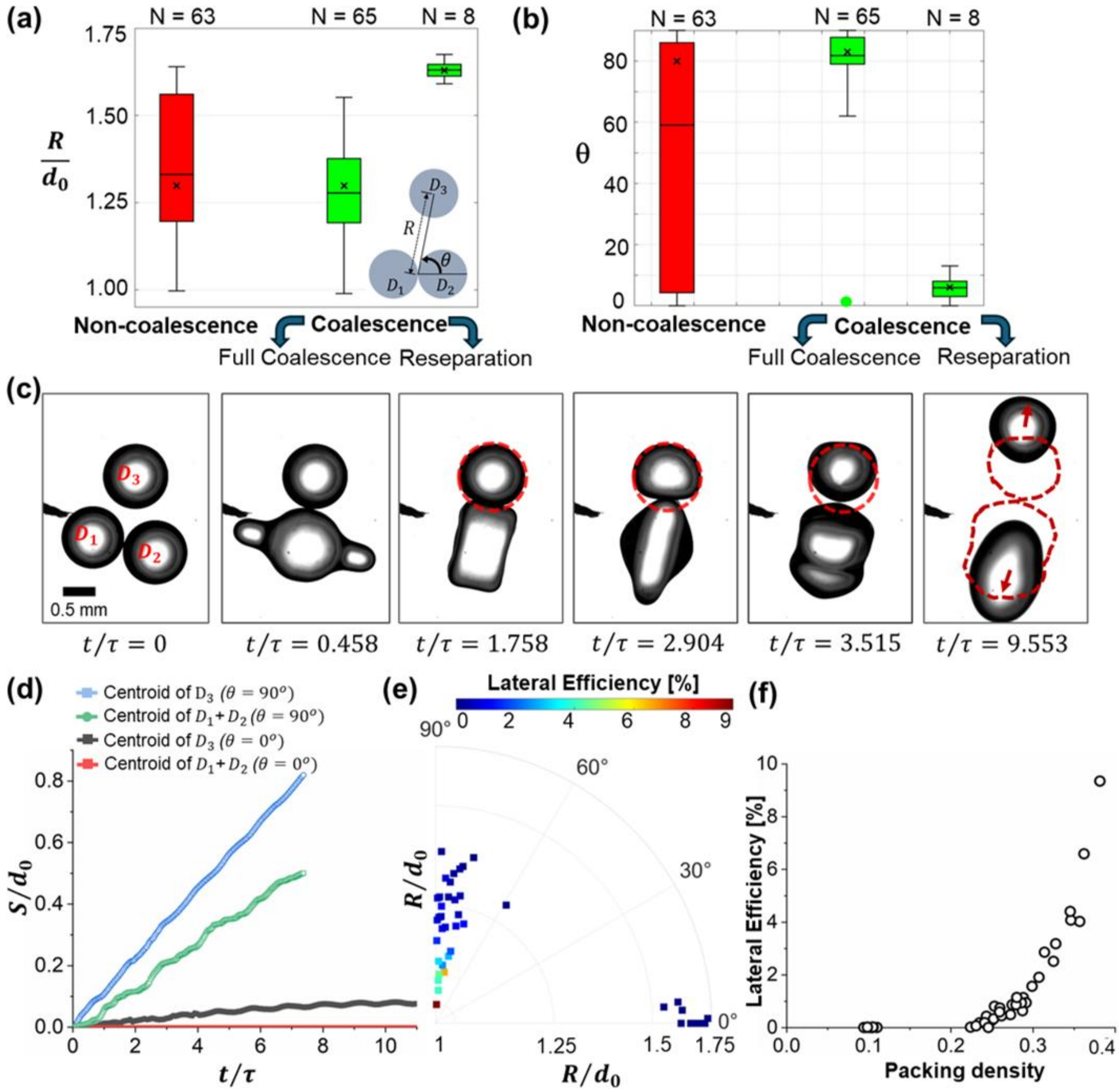

*Figure 5: The box plot for non-coalescence and coalescence for various $D_3$ locations defined in terms of (a) radial (R) and (b) angular (θ) coordinates as defined by the inset sketch. Here R is non-dimensionalized with droplet diameter $d_0$. N represents sample size. Coalescence events are separated into full coalescence and coalescence followed by reseparation. (c) Image sequence for a typical non-coalescence event in a cluster with staggered arrangement showing significant deformation and subsequent self-propulsion of droplet $D_3$. Red dashed contours mark droplet contours in the previous frame to highlight the deformation. Red arrows mark the direction of centroidal motion of droplets. (d) Temporal evolution of cumulative in-plane droplet displacement (S) for events shown in Fig. 5c and Fig. 2b. Here, $t/\tau = 0$ corresponds to the moment of loss of apparent contact (e.g. $t/\tau = 3.515$ in Fig. 5c). (e) Lateral efficiency of non-coalescing droplets for various cluster configurations. (f) Lateral efficiency as a function of packing density of the cluster.*

The kinetic energy for in-plane propulsion of droplets post a non-coalescence event arises from the excess surface energy available in the droplet cluster. We quantify this energy conversion by defining a lateral efficiency, $\eta_{lateral}$ as the ratio of the total lateral centroidal kinetic energy of all participating droplets, $KE_{T,lateral}$ post non-coalescence to the excess surface energy $\Delta E_s$ available. Here, $\Delta E_s$ is calculated as the difference between the final surface energy of the coalesced droplet $D_1 + D_2$ and the total surface energy of the individual droplets $D_1$ and $D_2$ prior to coalescence. Fig. 5e illustrates $\eta_{lateral}$ for non-coalescence events in various cluster configurations. The efficiency varies in the range ~ 1 – 9%, with an average value comparable to the efficiency of well-known coalescence-induced out-of-plane jumping of droplets on superhydrophobic surfaces.[35] The highest $\eta_{lateral}$ is achieved during non-coalescence in most closely packed staggered clusters (i.e. small $R/d_0$ and $\theta \approx 90°$). Overall, we can discern that the lateral efficiency increases with the packing density of cluster, as illustrated by the Fig. 5f. Here, we define the packing density of the cluster as the ratio of the total volume of the droplets in the cluster to the volume of the cylinder circumscribing the cluster.

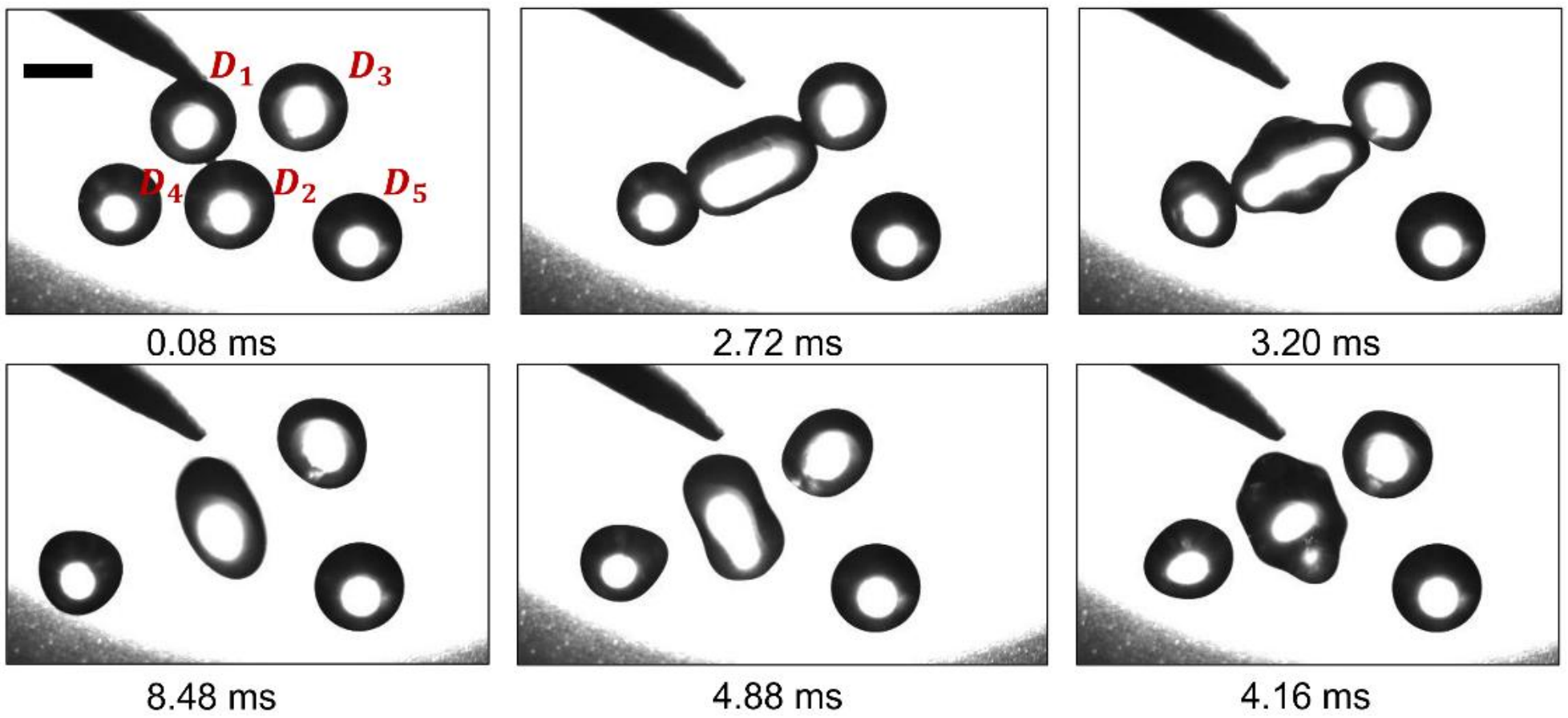

***Figure 6:*** *Non-coalescence of droplets in a cluster with more than 3 droplets. The scale bar in the first frame is 1mm. Here the actual number of participating droplets are 4 ($D_1$, $D_2$, $D_3$ and $D_4$). The droplet $D_5$ does not participate as it is out of reach. The formed interface, as a result of coalescence between $D_1$ and $D_2$, deforms the interface of droplets $D_3$ and $D_4$ simultaneously, resulting in higher total lateral centroidal kinetic energy.*

In addition, a higher lateral efficiency can be achieved if the number of participating droplets increases. Fig. 6 shows a closely packed cluster wherein $D_3$ and $D_4$ are located in the close vicinity of merging droplets $D_1$ and $D_2$. Here, non-coalescence is observed between the interface evolved from coalescence of $D_1$ and $D_2$, and the interfaces of droplets $D_3$ and $D_4$. The interface of $D_5$ does not coalesce as it is out of reach. Hence, the actual number of participating droplets in the cluster is 4. Here, we find $\eta_{lateral}$ is higher by ~ 2 percentage points compared to three-droplet cases in a similar arrangement due to more efficient utilization of surface energy in the deformation of multiple interfaces.

In addition to non-coalescence manifesting across a wide range of cluster configurations, we find that it can be seen for not only millimetric sessile droplet clusters, but also micrometric droplets formed through condensation. We observe this phenomenon during condensation on nanotextured superhydrophobic surfaces through in-situ observations of micrometric condensed droplets (refer to Section S5 for experimental details). Fig. 7a illustrates a typical non-coalescence event in a condensed droplet cluster, followed by in-plane self-propulsion of the participating droplets. The droplets participating in the event have diameters ranging from ~ 0.3 to 0.4 mm. For the event shown, the packing density and the lateral efficiency of the non-coalescence event is found to be 0.25 and ~0.6% respectively, which is comparable to the lateral efficiency observed in millimetric-sized droplets at a similar packing density.

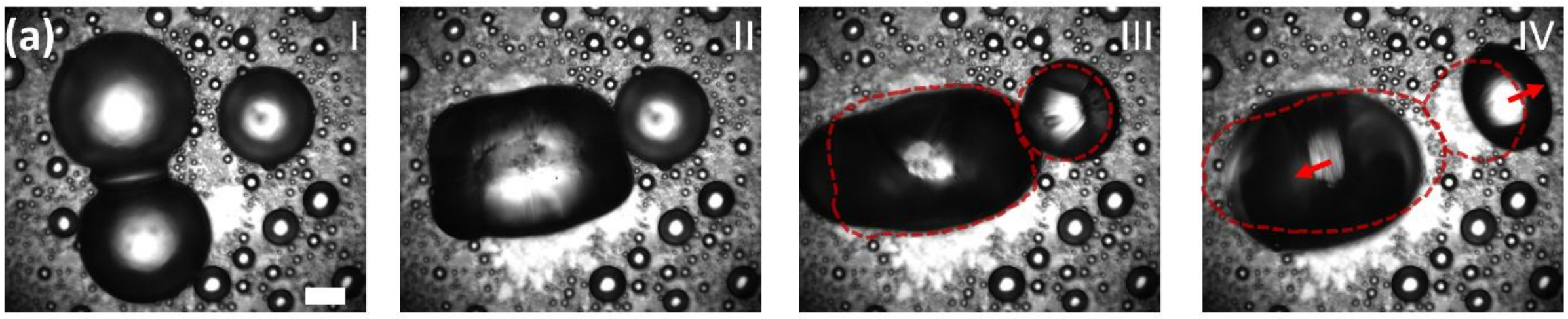

***Figure 7:*** *(a) Bouncing during condensation. Red dotted line represents the previous position of drops. Red arrows show the direction of centroidal motion of the individual drops. The scale bar represents 100 μm. The time difference between each image is ~ 1.4 ms. Subcooling is 10$^o$C and the relative humidity is 70%.*

## Conclusions

In this work, we have reported non-coalescence of sessile water droplet clusters on water repellant surfaces. For the first time, we observe non-coalescence between water droplets over a wide range of droplet sizes, ranging from millimetric sessile droplets to droplets as small as 200 μm. We find that such non-coalescence of droplet clusters manifests due to the presence of a stable air film that is formed mainly depending on the initial relative position of droplets in the cluster. We have investigated the air film dynamics numerically revealing that formation flat air film followed by a kink delays the air drainage between the interfaces which results in non-coalescence of droplets. We also find that non-coalescence in droplet clusters with close packing density can result in self-propulsion of droplets with significant in-plane kinetic energy, which is equally efficient as out-of-plane jumping. The lateral efficiency of such events can increase further with the number of participating droplets in the cluster. Finally, non-coalescence can manifest droplet clusters formed by condensation on nanostructured superhydrophobic surfaces, and the resulting self-propulsion of such small droplets throws light on a hitherto unknown pathway for spontaneous removal of droplets from the surface.

**Methods**

Substrate fabrication: Nanostructured superhydrophobic substrates were fabricated on glass and Aluminium substrates using spray coating and etching methods, respectively. Glass substrates were rendered superhydrophobic by spray coating of commercially available solution of silica nanoparticles, homogenously dispersed in isopropanol solution (Glaco Mirror coat "Zero". Soft 99 Co.).[21] Aluminium surfaces were made superhydrophobic by fabricating nanostructures through the Boehmite process and silanizing it with trichloro-1H,1H,2H,2H-perfluorodecylsilane (FDTS).[38,39] Here the surfaces were first subjected to the solvent cleaning process with acetone, IPA, and DI water, for 10 minutes each respectively under ultrasonication, followed by 2-minute cleaning with NaOH solution. Post-cleaning with DI water, the surface was then treated with the Boehmite procedure by placing the substrate in boiling DI water for 10 min.[39] The superhydrophobicity was achieved by treating the resulting nanotextured surface with a 1.43 mM solution of trichloro-1H,1H,2H,2H-perfluorodecylsilane (FDTS) from Sigma-Aldrich in n-hexane for 2 hours. This was followed by heating the surfaces on a hot plate at 120 °C for 45 minutes.

Wettability Characterizations: We measured apparent advancing ($\theta_a$) and receding contact angles ($\theta_r$) for all the surfaces using a contact angle meter setup (Holmarc Opto-Mechatronics Ltd, Model No- HO-IAD-CAM-01A). For superhydrophobic glass substrate, the $\theta_a$ and contact angle hysteresis ($\theta_a - \theta_r$) were found to be $162 \pm 1.6^0$ and $3.6 \pm 1.2^0$ respectively. Corresponding values for nanostructured superhydrophobic aluminum substrate were $161 \pm 1.1^0$ and $3.2 \pm 1.6^0$ respectively.

Experimental Methodology: Sessile droplet coalescence experiments were performed by carefully dispensing the required number of microliter-sized droplets on the prepared nanotextured superhydrophobic substrate. One droplet was slowly moved and brought in contact with the another droplet using a superhydrophobic wooden tip to trigger the initial coalescence. Images were recorded at or above 10000 frames per second, in side and top view depending on the droplet arrangement, using high-speed cameras (See Section S1 for more detail).

Numerical Analysis: The dynamics of the interstitial air film during non-coalescence of droplet clusters was investigated numerically by adopting the Volume-of-Fluid (VOF) method using ANSYS Fluent. We performed direct numerical simulations to model the evolution of the air film and dynamics of air drainage. Since the air film is much thinner compared to the droplet size, we optimized the computational effort by focusing on the non-coalescence in a droplet cluster with in-line arrangement of droplets shown in Fig 2. The inherent morphology and symmetry of the process enabled us to define the computational domain as shown in Fig. 3a (Refer to Section S3 for further details).

Calculation of air drainage velocity: The air drainage velocity was calculated as the magnitude of the resultant velocity in the radial (r) and axial direction (z) directions, given by $\sqrt{v_r^2 + v_z^2}$ . Here $v_r$ represents the air velocity in the radial direction and $v_z$ represents the air velocity in axial direction, relative to the axial motion of the interface formed by the merged droplet $D_1 + D_2$.

**Acknowledgements**

C.S.S. acknowledges the Indian Institute of Technology Ropar and Ministry of Human Resource Development, Government of India for partially funding this work through ISIRD (Grant No. 9-388/2018/IITRPR/3335) and STARS (Grant No. STARS-2/2023-0812) schemes, respectively. C.W.E.L. acknowledges support from the Croucher Foundation

**Author Contributions**

C.S.S. conceived and supervised the research and arranged the funding. C.S.S. and G.C.P. designed the experiments. G.C.P. fabricated the samples, performed the experiments, and conducted the numerical simulations. G.C.P., C.W.E.L., and C.S.S. analyzed the results. All authors contributed to writing the manuscript**.**

**Competing interest**

All authors declare they have no competing interests.

**Additional Information**

Supplementary materials are provided separately with this manuscript.

**Supplemental Material**

# Non-coalescence and in-plane momentum generation in sessile droplet clusters

Gopal Chandra Pal[†], Cheuk Wing Edmond Lam[‡], Chander Shekhar Sharma[†]*

[†]Thermofluidics Research Laboratory, Department of Mechanical Engineering, Indian Institute of Technology Ropar, Rupnagar, Punjab 140 001, India
[‡]Department of Mechanical Engineering, Massachusetts Institute of Technology, Cambridge, Massachusetts 02139, USA
*Email: chander.sharma@iitrpr.ac.in, Ph: +91-1881-232358

**Contents**

## S1. Experimental details for sessile droplet coalescence on superhydrophobic surfaces

We use both spray-coated and etched nanotextured superhydrophobic substrates for the coalescence of sessile droplets in the cluster. All the experiments are carried out on an actively vibration-isolated optical table (Daeil systems) to nullify the effect of outside vibrations on the experiments. The substrate is grounded to avoid any effect of static charge on the process (Fig. S1).[1] We use superhydrophobic micropipette tips developed in-house to dispense the droplets on the substrate. The superhydrophobic micropipette tips are fabricated by multiple cycles of dip coating in a nanoparticle-polymer solution followed by heating in the oven at 80 ºC. The polymer solution is prepared by mixing 17 mL of acetone (Sigma Aldrich), 3 mL of Capstone ST 200 (DuPont Fluoropolymer Solutions) and one gram of fumed hydrophobic silica nanoparticles (Evonik).[2]

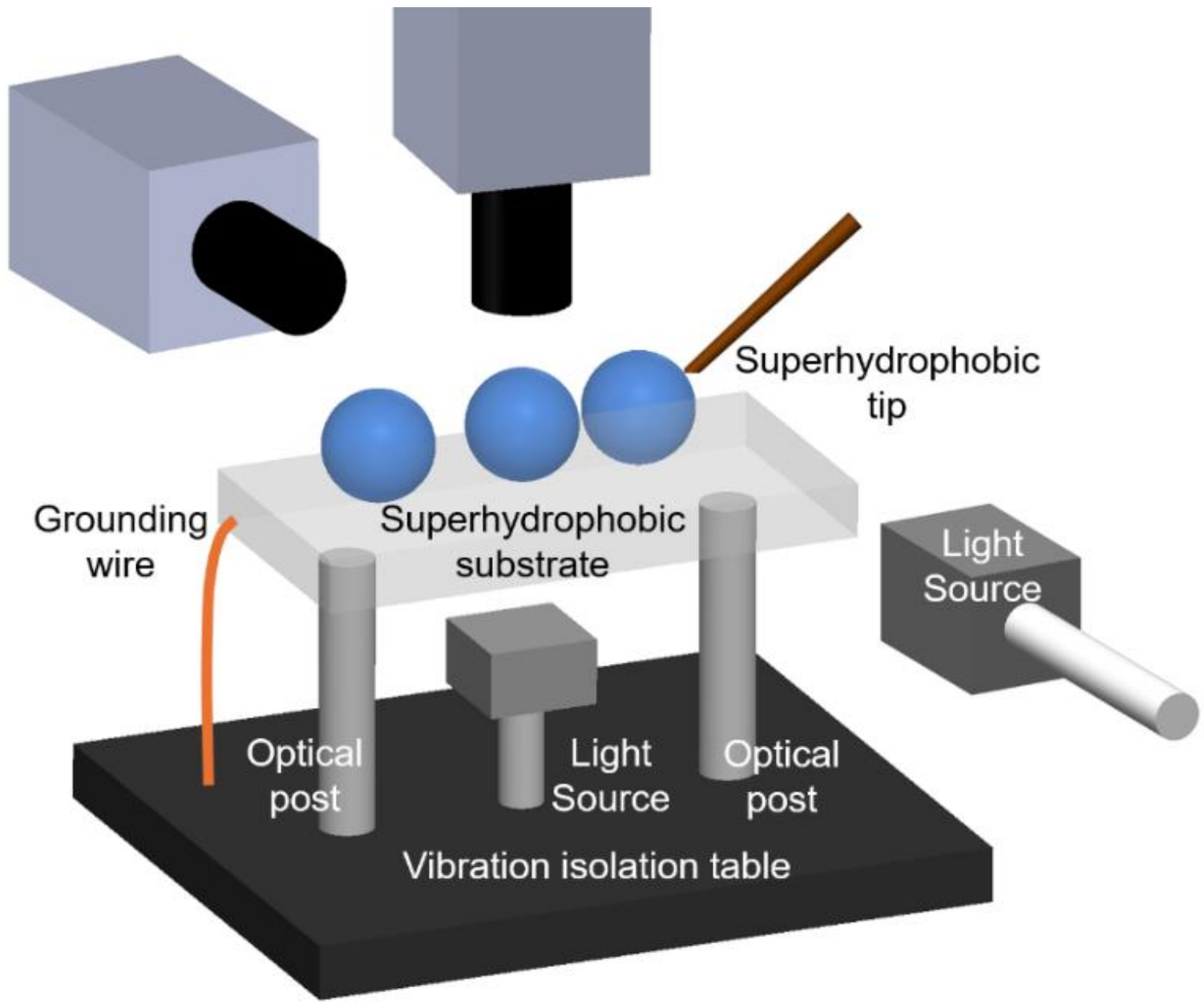

***Figure S1**: Experimental setup for coalescence-induced droplet jumping.*

In each experiment, the required number of microliter-sized droplets are first carefully dispensed on the substrate. Subsequently, one droplet is slowly moved and brought in contact with the another droplet using a superhydrophobic wooden tip to trigger the initial coalescence. We ensure, through image analysis, that any kinetic energy imparted to the droplet in this process is minimal (< ~5% of the excess surface energy of the droplet)[3]. Images are recorded at or above 10000 frames per second using Photron Fastcam SA4 high-speed camera for side view and high-speed camera (Photron Fastcam SA4) for top view as shown in Fig. S1. After the experiments, the images are analyzed for the temporal evolution of the events using Fiji ImageJ.

## S2. Droplet coalescence followed by reseparation

During coalescence of a sessile droplet cluster in an in-line arrangement with large initial gap between droplets $D_2$ and $D_3$, the evolving interface of the merged droplet $D_1 + D_2$ triggers coalescence with the third droplet $D_3$, but the coalescence does not proceed to completion and is followed by reseparation (See Fig. S2). In this process, a small daughter droplet is also generated, moving away at a velocity up to ~$0.4v_{ic}$, where $v_{ic}$ is the inertial-capillary velocity. The inertial capillary velocity is defined as: $v_{ic} = \sqrt{\frac{8\sigma}{\rho d_0}}$, where $\sigma$ , $\rho$ and $d_0$ are surface tension, density, and initial diameter of droplet respectively.[4,5] Reseparation manifests when the gap is nearly the maximum feasible gap for which the interface of $D_1$+$D_2$ can reach that of $D_3$ $\left({}^{l}/_{d} \gtrsim 0.14\right)$. At such large ${}^{l}/_{d}$ ratio, the interface of $D_1 + D_2$ starts to retract almost immediately after the initiation of coalescence. Consequently, the retraction rate dominates over the rate of coalescence neck growth between the evolving interface and $D_3$, resulting in reseparation of the two interfaces.

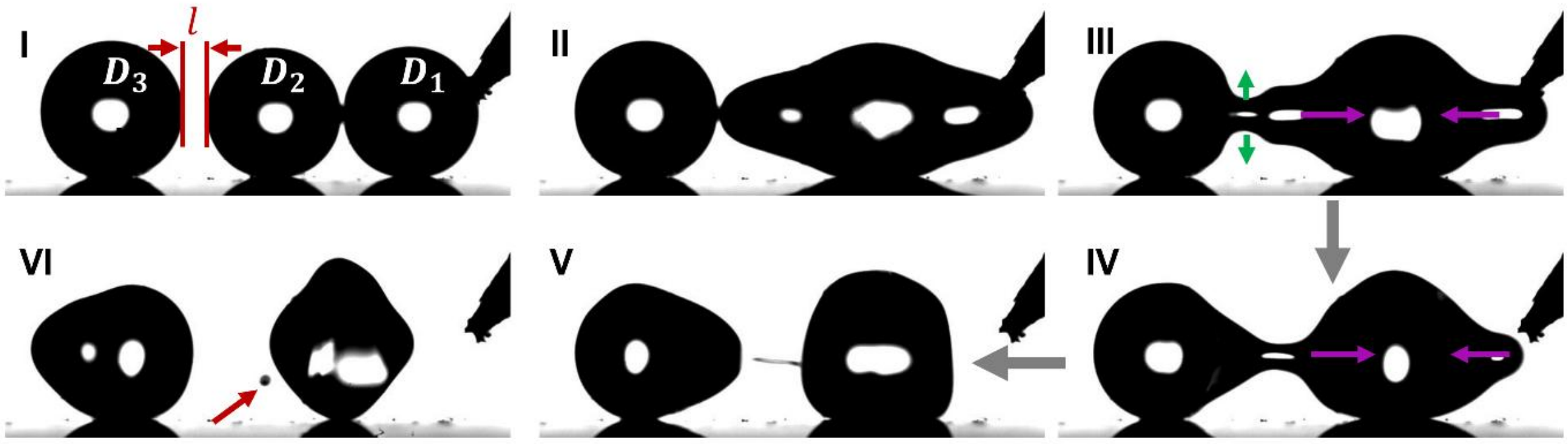

***Figure S2:*** *Coalescence followed by reseparation of droplets during the coalescence of a droplet cluster arranged in an in-line arrangement. The image in frame III shows neck growth (green arrow) between $D_2$ and $D_3$ and retraction of the evolving interface (purple arrow) of merged droplet $D_1 + D_2$. Reseparation and formation of daughter droplet (marked by red arrow) are shown in frames V and VI, respectively.*

## S3. Details of numerical modelling for the study of air film dynamics

We adopt Volume-of-Fluid (VOF) method-based Computational Fluid Dynamics (CFD) approach to model the non-coalescence of droplet clusters on superhydrophobic surface using ANSYS Fluent.[6] In a real scenario, coalescence or non-coalescence of droplets after apparent contact depends on molecular forces, which cannot be directly accounted for in the CFD simulations. Hence, the non-coalescence of droplets in literature is modelled using two main approaches: (a) multiscale approach, where multiphase simulation is combined with Reynolds thin film air drainage model.[7,8] The outcome of the coalescing event depends on the critical film thickness. (b) Geometric VOF method with distinct multi markers for different droplets and adaptive meshing.[9,10] In our numerical modelling, we have used the latter method, however, with a single marker for all the droplets, and an adaptive meshing strategy to resolve the morphology of air film and flow of air in the film.

We have chosen an incompressible, laminar flow model for our simulation. In this method, the following governing equations are solved for mass and momentum conservation of both liquid and gas phases:

$$\frac{\partial \rho}{\partial t} + \boldsymbol{\nabla} \cdot (\rho \boldsymbol{u}) = 0 \qquad (3.1)$$

$$\frac{\partial (\rho \boldsymbol{u})}{\partial t} + \boldsymbol{\nabla} \cdot (\rho \boldsymbol{u} \otimes \boldsymbol{u}) = -\boldsymbol{\nabla} p + \mu \nabla^2 \boldsymbol{u} + \boldsymbol{F}_{st} + \rho \boldsymbol{g} \qquad (3.2)$$

In the above equation, $\boldsymbol{u}$, $p$, $\rho$, $\mu$, and $t$ represents the velocity of mixture, pressure, density, viscosity, and time, respectively. The apparent viscosity $\mu$ and density $\rho$ in each cell is calculated by using the following equations:

$$\mu(x,t) = \mu_g + (\mu_l - \mu_g)\alpha \qquad (3.3)$$

$$\rho(x,t) = \rho_g + (\rho_l - \rho_g)\alpha \qquad (3.4)$$

Here, $\rho_l$ and $\rho_g$, $\mu_l$ and $\mu_g$ are the density and the viscosity of the liquid and gas, respectively. The volume fraction α represents the ratio of cell volume occupied by the liquid to the total volume of the control cell. The numerical value of $\alpha$ in a cell lies between 0 and 1, where $\alpha = 0$ in a cell indicates that the cell is filled with gas while $\alpha = 1$ indicates the cell is filled with liquid. Thus $\alpha$ tracks the evolution of the two immiscible phases where the interface between the two phases lies in the cells with $\alpha$ value lying between 0 and 1. The solver solves for the value of α in each cell based on the following transport equation.

$$\frac{\partial \alpha}{\partial t} + u \cdot \boldsymbol{\nabla} \alpha = 0 \qquad (3.5)$$

The volumetric force $F_{st}$ in equation (3.2) represents the effect of surface tension at the interface. Here, the Continuum Surface Force model[11] is used to capture the surface tension force as a volumetric force as per the following equation:

$$\boldsymbol{F}_{st} = \frac{\gamma_{lv}\rho\kappa\boldsymbol{\nabla}\alpha}{\frac{1}{2}(\rho_l+\rho_g)} \quad (3.6)$$

Here, the $\gamma$ and $\kappa$ represent the interfacial surface tension and the curvature of the interface, respectively. The surface curvature is calculated as $\kappa = -\boldsymbol{\nabla}\cdot(\boldsymbol{n})$ where $\boldsymbol{n}$ is defined as a unit normal vector given by $\boldsymbol{n} = \left(\frac{\boldsymbol{\nabla}\alpha}{|\boldsymbol{\nabla}\alpha|}\right)$.

In order to optimize the computational effort, we focus on non-coalescence in the in-line arrangement of droplets as shown in Fig. 2 (main paper) and consider the following salient aspects of this event: (a) The effect of the superhydrophobic substrate is negligible due to large static contact angle and low contact angle hysteresis.[5] (b) Further, for in-line droplet cluster, the evolving interface of merged droplets $D_1$ and $D_2$ reaches and retraces back from $D_3$ before the capillary ripple travelling on the interface of merged droplet is reflected from the substrate. Thus the overall process can be modelled using two-dimensional axisymmetric simulations wherein the axis of symmetry is chosen as shown in Fig. S3[12]. (c) Further, from the experimental results, we found that the placement of droplets on both sides or only on one side of the coalescing droplets $D_1$ and $D_2$ does not alter the outcome (coalescence/non-coalescence) of the process. Hence, we can further reduce the computational cost by assuming a symmetric boundary condition and thus solving for only one of the coalescing droplets $D_1$ and $D_2$. Based on the above discussion, the chosen domain and boundary conditions for the numerical simulation is shown in Fig. S3a and Fig. 3 in the main paper.

During the sessile droplet coalescence experiments, droplet diameter is varied in the range between 0.6 mm and 2 mm. The results show that the outcome of coalescence with a change in droplet size remains the same when the gap between $D_2$ and $D_3$, $l$, is scaled with diameter of the droplets since the coalescence of water droplets in this range is dominated by inertial-capillary forces.[4,5] Based on these results, the computational effort is further optimized by choosing the size of the droplets $d_0$~100 μm for our simulation. The domain size chosen in our simulation is $4d_0\times2d_0$.[13] Here, we have three mesh sizes: coarse (125 cells per droplet radius) in zone 1, fine (450 cells per radius) in zone 2 and finest cells (1800 cells per radius) at the interface of droplets as shown in Fig. S3a. Numerical results for non-coalescence of the droplet cluster are validated against the experimental results by comparing the temporal evolution of the droplets morphology, as shown in Fig. S3b. The coalescence time and droplet shape from experiments are non-dimensionalized with inertial-capillary time and length scales, respectively, for consistent comparison.[4]

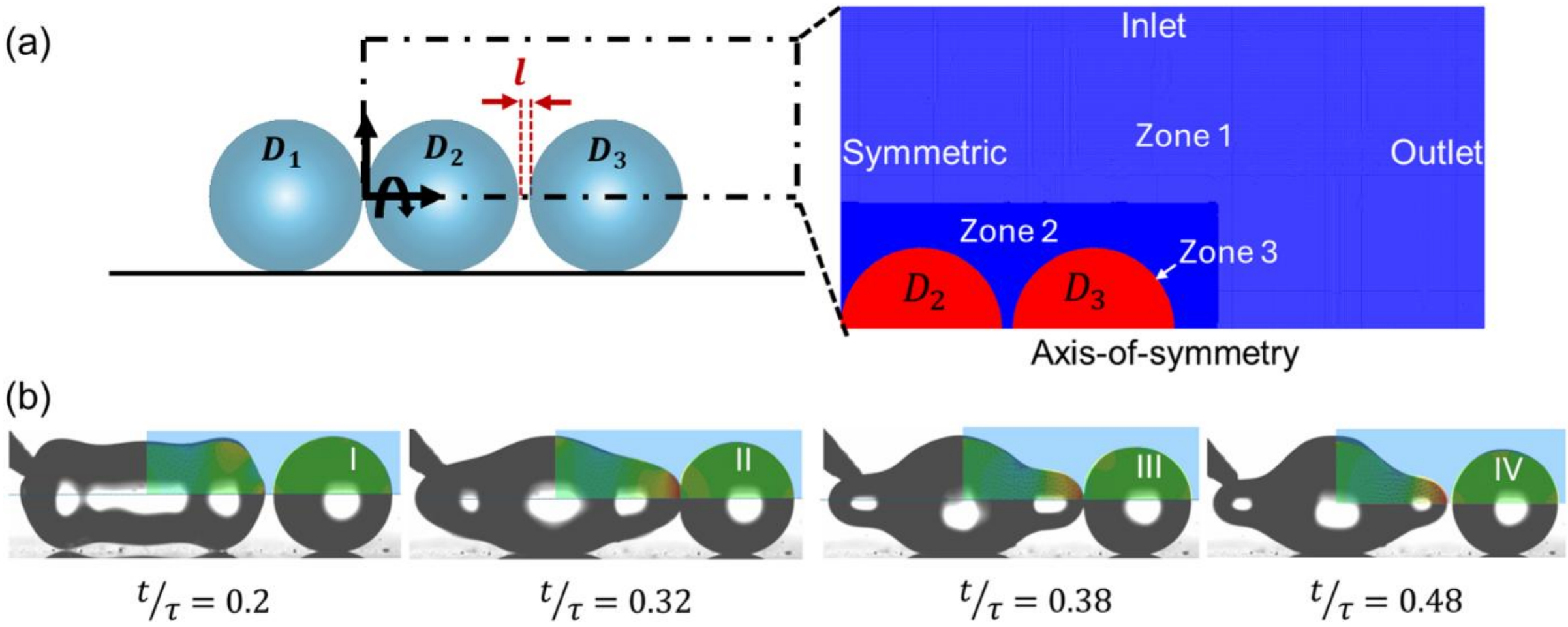

***Figure S3***. *Computational domain details and model validation. (a) Computational mesh and boundary conditions for the numerical modelling (b) Validation of numerical result with the experimental result at different time instances. Frame (I): Evolving interface of* $D_1+D_2$ *approaches* $D_3$*, Frames (II) and (III): Evolving interface stays in apparent contact and Frame (IV): the interface of* $D_1+D_2$ *retraces without triggering coalescence with* $D_3$*. The morphology of interfaces from numerical result is superimposed on top of experimental results. The droplet shapes and time from experiment were both scaled with inertial-capillary time and length scales for comparison with numerical results.*

**S4. Numerical methodology to obtain the region of coalescence for experiments**

In order to perform the experimental investigation in a systematic manner, we develop a simplified numerical methodology to estimate the possible region for droplet $D_3$ to be located within, such that the interface of merged droplets $D_1 + D_2$ can reach the interface of $D_3$. We assume that the presence of droplet $D_3$ does not influence the coalescence process of droplets $D_1$ and $D_2$ until the evolving interface reaches very close to $D_3$. First, the coalescence of two droplets on superhydrophobic substrates is numerically simulated by using the Volume-of-Fluid (VOF) method-based CFD approach described in Section S2. However, here, we solve the problem in 3D rather than using the axisymmetric 2D model. An incompressible, laminar flow model is chosen, and the contact angle on the surface is assumed to be $180^0$. This assumption is valid due to the high advancing contact angle and very low contact angle hysteresis on the fabricated surfaces.[14–16]

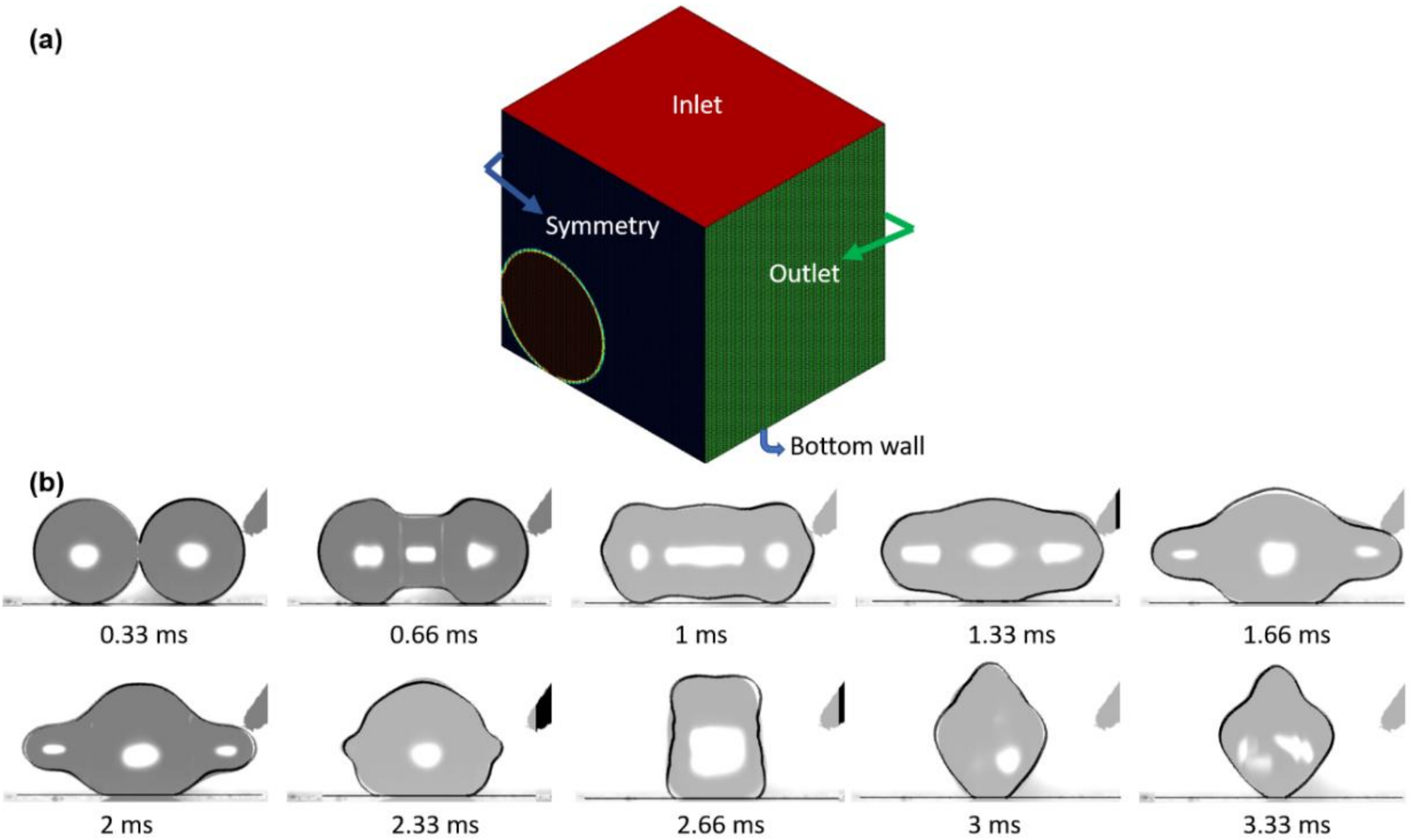

***Figure S4**: (a) Computational domain and the boundary conditions (b) Validation of numerical simulation with experimental results. The black line shows the droplet shape contour obtained from the numerical result and grey shaded shape shows the droplet shape obtained from the experimental result for coalescing droplets of diameter 1 mm.*

The domain size is chosen as $2d_0 \times 2d_0 \times 2d_0$ in our simulation, where $d_0$ is the droplet diameter. Two different mesh sizes are involved in the simulation: coarse mesh in the entire domain (40 cells per radius) and fine mesh at the interface (80 cells per radius). With the mesh density fixed, the numerical result for the coalescence of droplets with a diameter of 1 mm is validated with the corresponding experimental result on a superhydrophobic surface, as shown in Fig. S4a. The morphology of droplets from numerical simulation matches the experimental results quite well, as shown in Fig. S4b. The coordinates of air-water interface of the merging droplets $D_1$ and $D_2$ are extracted from the numerical results at different time steps and reconstructed in Matlab. Then, it is evaluated if the interface can reach that of droplet $D_3$ for a given centroidal position of $D_3$. The process is repeated for all $R$ and θ using an in-house Matlab code. Based on the mentioned analysis, we find the region of coalescence, i.e. possible locations for droplet $D_3$, as shown in Fig. S5.

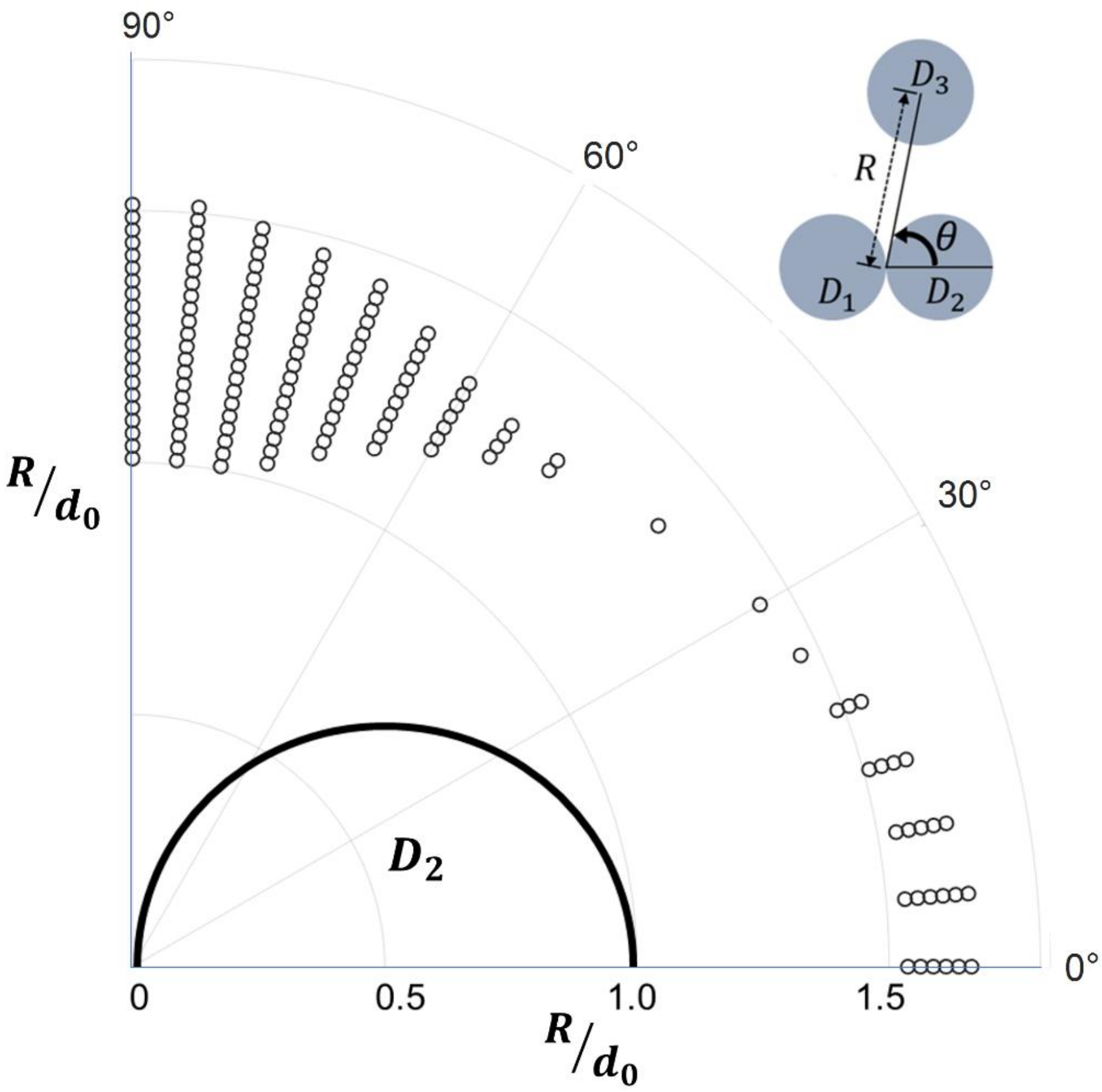

***Figure S5****: Polar plot indicating the position of $D_3$ for which the interface of merged droplets $D_1 + D_2$ can reach the interface of $D_3$. The small black circles in the figure depict the possible centroid positions of $D_3$, with respect to the coalescence point of $D_1$ and $D_2$. Size of all three droplets is assumed to be same i.e $d_0$. Inset sketch depicts the definition of $R$ and $\theta$.*

**S5. Experimental details for coalescence of droplets during condensation**

We perform in-situ condensation observations on a nanostructured superhydrophobic surface, by using a combination of high-speed imaging and optical microscopy, to observe the coalescence of small condensate droplets. The setup is shown in Fig. S6. Here, the substrate temperature is controlled using a recirculating chiller connected to the cold plate. The humidity is maintained inside the chamber by using a custom humidity setup, as depicted. Substrate contact with the cold plate is maintained using thermal paste (Arctic Silver). The images are recorded at frame rate ≥ 10000 fps. We also record humidity, ambient temperature, and substrate temperature inside the micro-condensation chamber. A subcooling of 10 ℃ and relative humidity of 70% are maintained in our experiments.

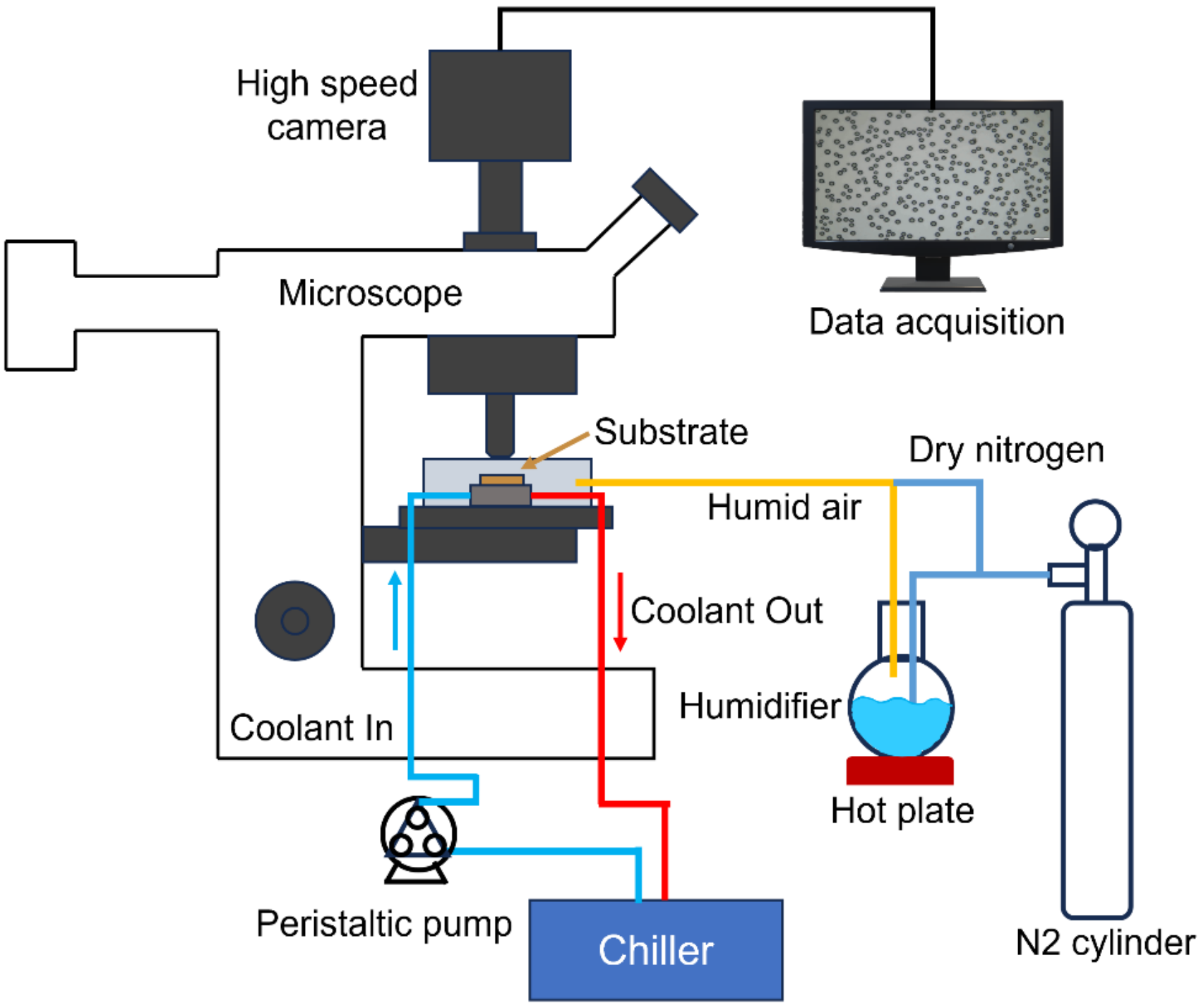

***Figure S6***: *Condensation setup*

**Supplementary References:**